\documentstyle[preprint,aps]{revtex}
\newcommand{\tQ}{\tilde Q}
\newcommand{\ra}{\rightarrow}
\newcommand{\be}{\begin{equation}}
\newcommand{\ee}{\end{equation}}
\begin{document}

{\tighten
\preprint{\vbox{\hbox{CALT-68-2086}
                \hbox{hep-th/9611049} 
		\hbox{\footnotesize DOE RESEARCH AND}
		\hbox{\footnotesize DEVELOPMENT REPORT} }}
 
\title{The Coulomb branch of N=1 supersymmetric gauge theory
with adjoint and fundamental matter \footnote{%
Work supported in part by the U.S.\ Dept.\ of Energy under Grant no.\
DE-FG03-92-ER~40701.} }
 
\author{Anton Kapustin}
 
\address{California Institute of Technology, Pasadena, CA 91125}

\maketitle 

\begin{abstract}
We consider $N=1$ $SU(N_c)$ gauge theory with an adjoint matter field
$\Phi$, $N_f$ flavors of fundamentals $Q$ and antifundamentals $\tQ$,
and tree-level superpotential of the form $\tQ\Phi^l Q$. This
superpotential is relevant or marginal for $lN_f\leq 2N_c$. The theory
has a Coulomb branch which is not lifted by quantum corrections.
We find the exact effective gauge coupling on the Coulomb branch 
in terms of a family of hyperelliptic curves, thus providing a generalization
of known results about $N=2$ SUSY QCD to $N=1$ context. The Coulomb branch
has singular points at which mutually nonlocal dyons become massless. These
singularities
presumably correspond to new $N=1$ superconformal fixed points. We discuss
them in some detail for $N_c=2, N_f=1$.
\end{abstract}
}

\newpage

\section{Introduction and Summary}\label{intro}
The phenomenon of electric-magnetic duality in supersymmetric gauge theories has
attracted a lot of attention recently. N. Seiberg presented strong evidence
that certain $N=1$ SUSY gauge theories flow to the same theory in the
 infrared~\cite{Seiberg}.
Soon afterwards many new examples of such ``duality'' were found, and it became
clear that this phenomenon is generic. However, the precise reason why the
dual gauge theories are so common is not understood, nor is it clear which
theories actually admit a dual description. A particularly mysterious case
is that of gauge theories with both fundamental and tensor matter and no
tree-level superpotential. Here we will focus on
$SU(N_c)$ gauge theories with one adjoint superfield $\Phi$, $N_f$
 fundamentals $Q$
and $N_f$ antifundamentals $\tQ$. A possible line of attack would be to
deform the action by adding a suitable relevant or marginal
superpotential, so that the moduli space be truncated and the theory become
 manageable. Kutasov and collaborators argued~\cite{K1,K2,K3}
that the operator ${\rm Tr}\Phi^{k+1}$
is relevant in the infrared for all $k<N_c$ and sufficiently small $N_f$,
and constructed the dual descriptions of the infrared physics in terms
of $SU(kN_f-N_c)$ gauge theory. In this letter we consider a different
 deformation, namely by the operator $\tQ\Phi^lQ$. It turns out that this
operator is relevant for $l<2N_c/N_f$ and marginal for $l=2N_c/N_f$.
The case $l=1$ is essentially the $N=2$ QCD studied by Seiberg, Witten and 
others. For $l>1$ we can only have $N=1$ SUSY, and therefore these theories
provide an $N=1$ generalization of Seiberg-Witten theories. Classically,
the theories we consider have Coulomb and Higgs branches. We focus
on the Coulomb branch, because unlike the $N=2$ case the Higgs
branch may be lifted by quantum corrections.

Our results can be summarized as follows:
 the matrix of the gauge couplings $\tau$ can be
 thought of as a normalized period matrix of a genus $N_c-1$ algebraic curve.
 Because of the electric-magnetic
 duality $\tau$ is not uniquely defined at any given point in the moduli space,
 but the curve itself is. We find these curves for arbitrary
superpotential of the form 
\be\label{super}
W=\sum_{i=0}^lh_{i\beta}^{\,\,\alpha}\,\tQ_{\alpha}\Phi^i Q^{\beta}
\ee
and any $N_f$ and $N_c$.
The family of curves we obtain is isomorphic to that describing $N=2$ SUSY QCD
 with $lN_f$ flavors. Thus at nonsingular points of the Coulomb branch the $N=1$
 theory with the supepotential eq.~(\ref{super}) flows to the same infrared 
theory as $N=2$ SUSY QCD with $lN_f$ flavors.
We can say that these two theories are dual to each other.\footnote{
 At the microscopic level they have different flavor symmetry groups, but in
the infrared, away from the singularities, the massless particles carry no
 flavor quantum numbers, and therefore the flavor symmetry is effectively
absent at low energies.}

The curves become singular at the submanifolds of the moduli space where
charged particles become massless. Computing the monodromies associated with
these singularities allows one to determine their electric and magnetic
 charges.
In general both electric and magnetic charges are nonzero, and we will refer
to such particles as dyons. If all charged massless particles are mutually
local, then the low-energy theory becomes a free field theory 
after an appropriate duality transformation. 
One can also tune the moduli and/or the superpotential so that several
 singularities corresponding to mutually nonlocal dyons collide. Analogy with
 Ref.~\cite{AD} suggests that at such points in the moduli space the
theory flows to an interacting $N=1$ superconformal field theory. These
superconformal theories are manifestly inequivalent to fixed points of
$N=1$ SUSY QCD discussed in Ref.~\cite{Seiberg}
(they have much smaller flavor symmetry groups). 

Though the singularity structure and the monodromies of the curves are the same
as for $N=2$ SUSY QCD with $lN_f$ flavors, the interpretation of the singular
points appears to be different. 
For example, $N=2$ SUSY QCD has points in the moduli space with $lN_f$ massless
 quarks and $SU(lN_f)$ flavor symmetry. In the theory with superpotential eq.~(\ref{super}), the same points
 have only $N_f$ charged quarks and therefore a smaller symmetry group.
 Similarly, the points corresponding to interacting superconformal field
theories presumably have only $N=1$ supersymmetry and are not isomorphic
to $N=2$ superconformal theories of Ref.~\cite{APSW}.
 As an example, we consider the case of $N_c=2$ for arbitrary relevant
 superpotential of the form eq.~(\ref{super}) and discuss the
 Argyres-Douglas points in the moduli space. 

\section{Generalities}\label{general}

Consider first the N=1 $SU(N_c)$ gauge theory with $N_f$ fundamentals $Q$,
$N_f$ antifundamentals $\tQ$, an adjoint field $\Phi$ and no superpotential.
It is very likely that this theory is in the nonabelian Coulomb phase at the origin of the moduli space~\cite{K1,K2,K3}.
 The anomalous dimensions of the fields cannot be
computed, but the vanishing of the beta-function~\cite{SV}
imposes one condition on them:
\be\label{gammas}
N_f\gamma_Q+N_c\gamma_\Phi=N_f-2N_c.
\ee
The dimension of the operator 
\be
M_{l\alpha}^{\,\,\beta}=\tQ_\alpha\Phi^lQ^\beta
\ee
 is given by
\be
{\rm dim}\,M_l=3-\frac{(2N_c-lN_f)(\gamma_\Phi+2)}{2N_f}.
\ee
Since $\gamma_\Phi+2$ is the dimension of a gauge invariant operator
 ${\rm Tr}\Phi^2$,
it is positive by unitarity. Thus we conclude that $M_l$ is a relevant 
operator for $lN_f<2N_c$, marginal for $lN_f=2N_c$, and irrelevant otherwise.
Classically, the chiral ring is truncated for $l<N_c$. Thus we can have both
the situation where the chiral ring is truncated classically but the operator
is irrelevant quantum-mechanically, and the situation where the classical
chiral ring is not truncated at all, but the operator is still relevant.
Note also that here we are in a better position than in the case of the 
${\rm Tr}\Phi^{k+1}$ perturbation, since we can determine exactly when our
perturbation is relevant.

Interestingly, when $lN_f=2N_c$ the operator $M_{l\alpha}^{\,\,\alpha}$ is not
 just marginal, it is $exactly$ marginal. To see this we note that such a
 perturbation preserves enough symmetry for
all $Q$'s and $\tQ$'s to have the same dimension. Then it is an easy matter
to check that the beta-functions for the coupling $h_l$ corresponding to 
$M_{l\alpha}^{\,\,\alpha}$ and for the gauge coupling $g$ are proportional,
 and thus in the $g-h_l$
plane there must be a line emanating from the fixed point $(g^{*},0)$ on
which both beta-functions vanish. For $l=1$ the line is actually just $g=h_1$. Moreover, if one sets $g=h_1$ from the very beginning, the theory
has $N=2$ SUSY and is scale-invariant. For $l>1$ the full theory cannot be
 scale invariant, since eq. (\ref{gammas}) requires
\be
2\gamma_Q+l\gamma_\Phi=2(1-l),
\ee  
and thus both anomalous dimensions cannot vanish simultaneously. Actually,
we know that for $lN_f=2N_c$ and $l>1$ the beta-function for the Wilsonian
 coupling is nonzero, and thus the theory generates a strong coupling 
scale $\Lambda$ (by definition, this is the scale at which the Wilsonian 
gauge coupling blows up in perturbation theory).

A remark is in order. For $l>1$ and $N_f<2N_c$ adding $M_l$ to the Lagrangian
 renders the theory nonrenormalizable. 
Therefore one should regard the perturbed theory
as an effective field theory valid below some ultraviolet cutoff $M$ which
is of the order $h_l^{-1/(l-1)}$. We will assume in what follows that the
strong coupling scale $\Lambda$ is much smaller than $M$.

We will consider adding the tree-level superpotential of the form
\be
W=\sum_{i=0}^l h_i M_i.
\ee
The coefficients $h_i$ transform in the $({\bf \overline{N}_f},{\bf N_f})$ 
of the flavor group $SU(N_f)_L\times SU(N_f)_R$.
Classically, the theory has a moduli space of vacua which has both Coulomb
and Higgs branches. We have nothing to say here about the latter
(it may not even exist in quantum theory). But it is easy to see that
the former cannot be lifted by quantum effects. Indeed, the theory
under consideration has an anomaly-free R-symmetry with
 $R_\Phi=0, R_Q=R_{\tQ} = 1$. 
Therefore any dynamically generated superpotential 
is quadratic in $Q,\tQ$ and cannot lift the Coulomb branch. In what follows,
when we mention the moduli space, we always mean the quantum moduli space
of the Coulomb branch.

\section{The curves}

The matrix of the Wilsonian $U(1)$ gauge couplings $\tau$ on the Coulomb branch
is a complex symmetric
$r\times r$ matrix, where $r=N_c-1$ is the rank of the group. It was explained
by Seiberg and Witten that any $Sp(2r,{\bf Z})/{\bf Z}_2$
 transformation of $\tau$
leaves physics invariant, and therefore $\tau$ may have nontrivial
 monodromies as one encircles singularities of the moduli space. 
Thus $\tau$ is best thought
of as a normalized period matrix of a family of genus $r$ algebraic curves. Our first task is to find
the appropriate family of the curves. The arguments are a straightforward
generalization of those in Ref.~\cite{APS}. We assume that the curve is
hyperelliptic, and consider first the case $l N_f=N_c$. The symmetries
and the requirement that the limit $\Lambda\ra 0$ give the classical results
constrain the curve to be of the form 
\be
y^2=P_{N_c}(x,\phi)^2-\Lambda^{2N_c-N_f} Q(x,\phi,h)-\Lambda^{4N_c-2N_f} R(x,\phi,h).
\ee
Here $P_{N_c}(x,\phi)=\prod_i (x-\phi_i)$, and $\phi_i, i=1,\ldots N_c$ are
 the eigenvalues of $\Phi$. Giving the first $k$ eigenvalues a vev much larger
than $\Lambda$ (but smaller than the ultraviolet cutoff $M$) breaks
the gauge group down to $SU(k)\times SU(N_c-k)\times U(1)$ at weak coupling.
We can also tune the coefficients $h_i, i=0,\ldots,l$ so that the low-energy
theory still has $N_f$ quarks in the $({\bf k, 1})$ representation of
the gauge group.
In this limit the curve should factorize appropriately. In fact, factorization 
requires that $R$ be identically zero and $Q$ be independent of $\phi_i$.
We conclude that the curve has the form
\be
y^2=P_{N_c}(x,\phi)^2-\Lambda^{2N_c-N_f}Q(x,h).
\ee
To determine the form of the polynomial $Q$ we give all the quarks
 large masses. The quark mass matrix is $h_0$, so the limit we wish
to  consider corresponds to the eigenvalues of $h_0$ being much larger than
$\Lambda$, though still smaller than $M$. In this regime the semiclassical
analysis is adequate, and we know that the curve must be singular whenever
one of the eigenvalues $\phi_i$ is the root of the polynomial
\be\label{Q}
{\rm det}\,\sum_{i=0}^l h_i x^i.
\ee
This requirement determines that $Q(x,h)$ must be given by eq. (\ref{Q}).

To obtain the curves for $lN_f\neq N_c$ we may either integrate out some
of the quarks by giving them large masses, or to break the gauge group at
weak coupling by large adjoint vevs. The result is that for any $N_f, N_c$ and
$l$ such that $lN_f<2N_c$ the curve is given by
\be\label{curve}
y^2=P_{N_c}(x,\phi)^2-\Lambda^{2N_c-N_f} {\rm det}\,\sum_{i=0}^l h_i x^i.
\ee
Note that the ``quantum piece'' of the curve has degree in $x$ smaller than
the classical piece if $lN_f<2N_c$, which is exactly the condition for the perturbation to be relevant! Thus the curve has genus $N_c-1$, as required,
for any $l$ for which the perturbation is relevant.

The case $lN_f=2N_c$ is somewhat special in that the perurbation by $M_l$
is marginal rather than relevant. The curve may depend then on the dimensionless
parameter $t=\Lambda^{2N_c-lN_f} {\rm det}\, h_l$. This is in agreement with the
discussion in the previous section where it was argued that for $lN_f=2N_c$
there exists an exactly marginal perturbation of the theory without
the tree-level superpotential. The preceding discussion then does not
determine the polynomial $Q$ uniquely. Instead, we only obtain that
it is given by eq. (\ref{Q}) with $h_i$'s replaced by
\be
{\tilde h}_i=h_i G_i(t),
\ee
where the functions $G_i(t)$ are analytic at $t=0$ and satisfy 
$G_i(0)=1$ for all $i$'s. One may regard this as some nonperturbative renormalization of the coefficients $h_i$. We cannot determine the exact form
of this renormalization, but this is not important, since the region
of interest $\Lambda \ll M$ coresponds to $t\ll 1$, and we can just approximate
$G_i$'s by 1. 
Note also that for $lN_f=2N_c$ the second term in eq. (\ref{curve})
is of the same degree in $x$ as the first term, and the curve still has the
right genus.

As a check on the curve, one can compute the semiclassical monodromies and
 compare them with those given by the curve. Here we will do this 
only for $N_c=2$. It easy to see that traversing a large circle in the complex 
 plane of $u=\frac{1}{2}{\rm Tr}\Phi^2$ induces the monodromy
 ${\cal M}_\infty=T^{lN_f/2-2}$.
(The power of $T$ may be half integer because we normalized the
charge of the quark to be $1/2$ rather than one. The resulting sign
ambiguity has no influence on $\tau$, since the duality group is
$SL(2,{\bf Z})/{\bf Z}_2$ rather than $SL(2,{\bf Z})$.)
On the other hand, at large $u$ all charged particles acquire large masses,
and the gauge coupling stays small all the way down to extreme infrared.
 Thus the perturbative computation is
adequate in this regime. In fact, there are no perturbative corrections to the
 (Wilsonian) gauge coupling beyond one loop~\cite{SV}. To do the one loop
 computation, we
 notice that the charged particles originating from the adjoint fields acquire
masses of the order $u^{1/2}$ from the $D$-terms, 
while the fundamental matter gets mass only from the $F$-terms.
 This ``$F$-term'' mass is of the order $h_l u^{l/2}\ll u^{1/2}$.
Thus first we must integrate out the adjoint matter and $W$'s, and then the
fundamental matter. The resulting expression for the gauge coupling
\be
\tau(u)=i\frac{4-lN_f}{4\pi}\log u +const
\ee
 produces the right monodromy. In particular, for $lN_f=4$, the 
effective gauge coupling does not depend on $u$ in perturbation theory.

One can easily see that the family of curves in eq.~(\ref{curve}) is identical
to the family of curves describing $N=2$ $SU(N_c)$ SUSY QCD with $lN_f$
fundamental hypermultiplets~\cite{APS}. Away from singular points the
infrared limit is just a free field theory of photons, neutral scalars
and their superpartners, and thus automatically has $N=2$ SUSY.
 We may say that our model is dual to $N=2$ SUSY QCD with $lN_f$ 
flavors. The operators parametrizing the flat directions on the moduli space,
 ${\rm Tr}\Phi^i, i=2,\ldots, N_c,$ map trivially between the two theories.

For $l=1$ the theory actually
flows to $N=2$ SUSY QCD everywhere in the moduli space, including
 singularities~\cite{LS}. For $l>1$ this is no longer true, as
illustrated in the next section, where we analyze in more detail the case of $N_c=2$.   

\section{Examples}
\subsection{$N_c=2, l=1$}\label{1}
The curve is given by
\be
y^2=(x^2-u)^2-\Lambda^{4-N_f} {\rm det}\,(h_0+h_1x).
\ee
This case has been discussed previously in Refs.~\cite{LS,K3}. It was shown in
\cite{LS} that for $N_f=2N_c$ the theory flows to $N=2$ SUSY QCD. Giving
quarks small masses does not destroy $N=2$ SUSY, and since we expect that
there is no phase transition between small and large masses, the theory flows
to $N=2$ SUSY QCD everywhere in the moduli space provided that $N_f\leq 2N_c$.
 
\subsection{$N_c=2,N_f=1,l=2$}\label{2}
The curve is given by
\be
y^2=(x^2-u)^2-\Lambda^3(h_0+h_1x+h_2x^2).
\ee
It is isomorphic to the curve of $N=2$ SU(2) SUSY QCD with two
 flavors of quarks,
and therefore we can take over the results of Ref.~\cite{APSW}. 
For generic values
of $h_0,\ldots,h_2$ there are two quark singularities, \footnote{The masses of
up and down quarks always vanish simultaneously, and when we refer to 
a quark, what we mean is the pair of up and down quarks.} 
one point where the (1,0) dyon becomes massless, and one point
where the (1,1) dyon becomes massless. If we set $h_1^2=4h_0h_2$, the quark
singularities collide. This does not mean, however, that there are two massless
quarks at this point. This is quite obvious for large $h_0$, where the 
quark singularity occurs in the semiclassical regime, and must be true
in the strong coupling regime by continuity. Rather, we still have
one massless quark, but its mass vanishes quadratically as one approaches
the singular point. 

Furthermore, one can make the quark singularity collide with one of the dyonic
 points by tuning $h_0$.
The result is a theory with one massless quark and one massless dyon.
Such a theory is intrinsically strongly interacting~\cite{AD},
 and presents a new 
example of an $N=1$ superconformal fixed point. It has no apparent
nonabelian global symmetries, and therefore it is inequivalent to
the fixed points of $N=1$ SUSY QCD~\cite{Seiberg}. 

We can also compute the scaling dimensions of relevant perturbations around
this point 
up to overall normalization. (Unlike Ref.~\cite{APSW}, we cannot
determine the normalization, because we do not have dyon mass formulas
or any information about the K\"ahler potential.) In fact, the answer can be
read off from section 3.2 of Ref.~\cite{APSW}:
\be
D(h_1):D(\delta u):D(h_1^2-4h_0h_2)=1:2:3.
\ee

\subsection{$N_c=2,N_f=1,l=3$}\label{3}
The curve is given by
\be
y^2=(x^2-u)^2-\Lambda^3(h_0+h_1x+h_2x^2+h_3x^3).
\ee
For generic values of $h_0,\ldots,h_3$ there are two dyonic singularities
and three quark singularities. Tuning the coefficients of the superpotential
one can collide all three quark singularities and produce a point $u=u_0$
with one massless quark. The mass of the quark vanishes like
$(u-u_0)^3$ as $u$ approaches $u_0$. By further tuning one can collide this point with one of the dyonic singularities, producing an 
interacting $N=1$ superconformal theory. It has no nonabelian flavor
 symmetries. There are four independent relevant perturbations away
from the superconformal point:
\be
 \delta u,\quad h_2,\quad 3h_1h_3-h_2^2,\quad 27h_0h_3^2-9h_1h_2h_3+2h_2^3.
\ee
Their respective scaling dimensions are in the
ratio $3:1:4:6$. Thus this superconformal theory is manifestly
 inequivalent to that discussed in the previous subsection. 

\subsection{$N_c=2,N_f=1,l=4$}\label{4}

The curve is given by
\be
y^2=(x^2-u)^2-\Lambda^3(h_0+h_1x+h_2x^2+h_3x^3+h_4x^4).
\ee
Now there are four quark singularities. By tuning the parameters of the
superpotential we may collide three of them with one of the dyonic
 singularities. This occurs at isolated points of the parameter space of the
theory and presumably produces the infrared fixed point discussed in the
 previous subsection. There are also submanifolds in the parameter space where only
two quark singularities coincide with the dyonic one. On these submanifolds
the theory flows to the fixed point of subsection~\ref{2}.

 If we set $h_0=h_1=h_2=h_3=0$, all six singularities collide at the
origin of the moduli space. As discussed in section~\ref{general},
the perturbation by the operator $Q\Phi^4Q$ is truly marginal for
 $N_c=2,N_f=1$. In other words, there is a line of interacting superconformal
 fixed points in the $g-h_4$ plane emanating from the point $(g^*,0)$, where
 $g^*$ is the value of the gauge coupling at the fixed point of
the theory without superpotential. Thus the point $u=0$ is naturally 
interpreted as a nonabelian Coulomb phase point. The relative dimensions
of the perturbations determined from the curve coincide with those obtained
from the microscopic $R$-charge assignments.\footnote{In this case there is
a one-parameter family of nonanomalous microscopic $R$-symmetries.}
This is in agreement with the assumption that one can describe the 
nonabelian Coulomb phase in terms of microscopic degrees of freedom. 

\acknowledgements
It is a pleasure to thank John Preskill, David Lowe and Eric Westphal
for helpful discussions and comments.
 I am especially grateful to Adam Leibovich for collaboration on some related
matters.

 \end{document}